\documentclass[preprints,communication,accept, pdftex,moreauthors]{Definitions/mdpi} 

\usepackage{amssymb, bm, tensor}
\usepackage{amsthm, thmtools}

\usepackage{microtype}

\usepackage{cancel}

\usepackage{enumitem}

\usepackage{graphicx}

\newcommand{\R}{\mathbb R}

\newcommand{\mnu}{{\mu\nu}}
\newcommand{\beq}{\begin{equation}}
\newcommand{\eeq}{\end{equation}}
\newcommand{\beqNo}{\begin{equation*}}
\newcommand{\eeqNo}{\end{equation*}}

\newcommand{\CR}{\mathcal R}

\newcommand{\p}{\partial}
\newcommand{\const}{\mbox{\rm const}}

\newcommand{\II}{{\rm II}}

\newcommand{\diff}{{\mathrm d}}
\newcommand{\diffb}{{\mathbf d}}

\newcommand{\ddt}[1]{\frac{{\mathrm d} #1}{{\mathrm d} t}}

\newcommand{\Rwarp}{\textbf{R}-Warp}
\newcommand{\Twarp}{\textbf{T}-Warp}
\newcommand{\Fwarp}{\textbf{R1}-Warp}
\newcommand{\natario}{Nat\'ario}

\Title{On Restrictions of Current Warp Drive Spacetimes and Immediate Possibilities of Improvement}

\TitleCitation{On Restrictions of Current Warp Drive Spacetimes and Immediate Possibilities of Improvement}

\Author{Hamed Barzegar $^{\dagger}$\orcidA{} and Thomas Buchert *$^{,\dagger}$\orcidB{}}

\AuthorNames{Hamed Barzegar and Thomas Buchert}

\AuthorCitation{Barzegar, H.; Buchert, T.}

\address [1]{Centre de Recherche Astrophysique de Lyon, CNRS UMR5574, Univ Lyon 1, Ens de Lyon, \linebreak F-69007 Lyon, France; hamed.barzegar@ens-lyon.fr}
\corres{\hangafter=1 \hangindent=1.05em \hspace{-0.82em} Correspondence: buchert@ens-lyon.fr} 

\firstnote{\hangafter=1 \hangindent=1.05em \hspace{-0.82em} These authors contributed equally to this work.}

\abstract{Looking at current proposals of so-called `warp drive spacetimes', they appear to employ General Relativity only at an elementary level.
A number of strong restrictions are imposed such as flow-orthogonality of the spacetime foliation, vanishing spatial Ricci tensor, and dimensionally reduced and coordinate-dependent velocity fields, to mention the main restrictions. We here provide a brief summary of our proposal of a general and covariant description of spatial motions within General Relativity, then discuss the restrictions that are employed in the majority of the current literature. That current warp drive models are discussed to be unphysical may not be surprising; they lack important ingredients such as covariantly non-vanishing spatial velocity, acceleration, vorticity together with curved space, and a warp mechanism.
}

\keyword{general relativity; covariant motion; warp drive spacetimes}

\begin{document}
%

\section{Brief Recap on the State-of-the-Art of Warp Drive~Spacetimes}
\label{recap}

The description of accelerated motions within Special Relativity (SR) enjoys popularity in lectures and exercises when applied to the idea of space travel (see, e.g.,~\cite{Gourgoulhon_2013_SRT, Mumford_2021_AMS, Rindler_1969_essential}). 
However, such travels described within SR face limitations due to the flatness of spacetime and lightcone barrier (see, e.g.,~\cite{2011_Geroch_FTL} for stimulating thoughts and discussions).

A smarter way to think of space travels was put forward by Alcubierre~\cite{Alcubierre_1994_warp} (see also~\cite{Alcubierre_2017_warp_basics}), 
motivated by notions of science fiction. Alcubierre aimed at exploiting the far-reaching possibilities offered by General Relativity (GR)\footnote{Another area of research within GR involves traversable wormholes, which occasionally exhibit similarities with the current 
warp drive models (see, e.g.,~\cite{1988_Thorne_Morris_Wormholes, 1996_Visser_LorentzianWormhole, 1995_Krasnikov_Hyperfast}).}. Since expansion or contraction of space does not fall under the special-relativistic limitation of the speed of light---although locally causality is respected---a hyperfast space travel seems to be possible by actively deforming the gravitational field that, according to Einstein, defines the spacetime geometry.
Often, the~recession velocity in a globally expanding cosmological model  is employed as an example where the speed of light can be exceeded, while a (comoving) observer in the Universe would be confined to the Hubble sphere (cf., e.g.,~\cite{1993_Ellis_Rothman_Horizons, 2003_Davis_Lineweaver_Misconceptions}).

Since Alcubierre's proposal, a~substantial body of literature has been developed that is essentially based on the space travel design introduced by Alcubierre and later Nat\'ario who dropped the restriction of a one-component coordinate velocity field (related to the shift vector field) and considered the possibility of vanishing expansion~\cite{Natario_2002_warp}. The~subject of spacetime engineering also caught the attention of researchers at space agencies realizing the potential of such theoretical ideas, e.g.,~\cite{White_2003_Engineering, White_2011_propulsion}. In~\cite{Visser_2021_generic_warp}, the~authors summarize most of the literature assigning the wording `generic warp drives' 
to these concepts. We refer to this paper for a comprehensive list of references, but~we do not comment on its contents here.
Contrary to the wording `generic', we here name those concepts `restricted warp drives' (\Rwarp{} for short), and~we explain how we can and should drop a number of highly restricting~assumptions.

In this paper, we concentrate on the covariant description of spatial motion in GR as opposed to the coordinate-dependent architecture of current warp drive proposals. We think that such a description is key to open the possibility of a future definition of what the idea behind warp drive means in terms of Einstein's equations. We remain within classical GR and also leave the description of extended `warp bubbles' and their motion to later works. 
We emphasize that current proposals are confined to describe motions within a given 3+1 
foliation of spacetime, and hence they are subject to the classical Cauchy problem of GR. The~idea of warp drive implies an active gravity control or spacetime engineering process that locally changes the spacetime structure during the travel. 
We think that there is a long way ahead to arrive at a physical definition of warp drive. In~particular, we notice that current proposals do not introduce a warp mechanism, mainly due to an externally given velocity. Despite this, we employ the wording `warp' to exhibit the aim of our path towards this concept. A~next step consists in establishing all evolution equations, constraints, and conservation laws governing the warp field,
linking all relevant fields to variables that could eventually be actively controlled by the~spaceship.

We organize this paper as follows. In~Section~\ref{covkinematics}, we first recall well-known geometrical concepts for the description of covariant spatial motions and their kinematics. We then put currently employed restrictions into perspective and explore properties of \Rwarp{} designs in Section~\ref{SR}. We contrast this restricted concept to our proposal of tilted warp drives (\Twarp{} for short) within the covariant setting in Section~\ref{S}, followed by a comparison of the two concepts in Section~\ref{comparison}. In~Section~\ref{outlook}, we lay down a strategy on how to improve on current warp drive realizations and discuss some perspectives.
This work is chiefly based on and inspired by formalism developed in the context of inhomogeneous cosmology~\cite{buchertetal:generalfluid}.

\textbf{Notations:} 
Throughout, bold letters are used for coordinate-free notation of  given vector and tensor fields whose components are given in local coordinate bases $\{\bm \diff x^\alpha, \bm \p_\alpha \}$ 
with $\left( x^\mu \right) = \left( t, x^i\right) \equiv \left( t, \bm x \right)$, where Greek indices denote the spacetime coordinates on $(\mathcal{M}, \bf g)$, the~spacetime manifold  equipped with a Lorentzian metric, whereas Latin indices stand for spatial coordinates.
We set the speed of light at $c = 1$.

\section{Covariant Spatial~Motion}
\label{covariant}
\unskip

\subsection{General Kinematics of Covariant Spatial~Motion}
\label{covkinematics}

The current literature on warp drive spacetimes is based on a 3+1 decomposition of spacetime, foliated into spatial leaves $\Sigma_t$ with the unit normal $\bm n$, i.e.,~$\mathcal{M} = \Sigma_t \times I$ with $t \in I \subset \R$.
The components of $\bm n$ and its metric dual $1$-form\footnote{A bar under the symbol denotes the metric dual $1$-form of a given vector field tangent to the given manifold.} are given by
\begin{equation}\label{eq:n_vec}
	\bm{n} = \frac{1}{N} \left( 1, - \bm{N} \right)
\,,
\quad
	\underline{\bm n} = - N \, ( 1, \bm 0 ) \, ,
\end{equation}
where $N$ is the lapse function and $\bm N = N^i \bm{\partial}_i$ is the shift vector (cf., e.g.,~\cite{buchertetal:generalfluid,smarr:foliation,gourg:foliation}). 

The bilinear form $\mathbf{h} = h_{\alpha \beta} \, \mathbf{d} x^\alpha \otimes \mathbf{d} x^\beta$ on $\Sigma_t$ can be used to project spacetime tensors onto the hypersurfaces of the foliation and has the following form and properties:
\begin{equation}\label{eq:proj_h}
\begin{aligned}
	h_{\mu \nu} &:= g_\mnu + n_\mu n_\nu
\, ,
\\
	h_{\alpha \mu} n^\alpha &= 0 
\, ,
\quad
	\tensor{h}{^\mu_\alpha} \tensor{h}{^\alpha_\nu} = \tensor{h}{^\mu_\nu} \, , \quad 
	h^{\alpha\beta} h_{\alpha\beta} = 3 \, , 
\end{aligned}
\end{equation}
defining the induced Riemannian metric $h_{ij}$ on the hypersurface $\Sigma_t$  whose inverse is denoted by $h^{i j}$ and~the projection operator $\tensor{h}{^\mu_\nu}$. 
Hence, one can decompose the four-dimen\-sional line element into 3+1 form ($N_i = h_{i j} N^j$) \cite{adm}:
\begin{equation}\label{eq: ADM metric}
\begin{aligned}  
    {\mathrm d}s^2 
	&= 
    g_{\alpha\beta} (t, {\bm x})\, {\mathrm d}x^\alpha {\mathrm d}x^\beta 
\\
    &= 
    - \left( N^2 - N^k N_k \right) {\mathrm d}t^2 + 2 N_i \, {\mathrm d}x^i \, {\mathrm d}t 
      + h_{ij} \, {\mathrm d}x^i {\mathrm d}x^j 
\, .
\end{aligned}
\end{equation}

In this paper, we consider a general relativistic fluid as our matter model, which implies the existence of a unique $4$-velocity $\bm u$.
The fluid $4$-velocity $\bm u$ can in general be decomposed as follows (cf., e.g.,~\cite{buchertetal:generalfluid,gourg:foliation, smarr:foliation}):
\begin{equation}\label{eq:four_vel}
\begin{aligned}
    {\bm u} 
    &= 
    \gamma ({\bm v}) \left( \bm n + \bm v \right) 
\ , \ \ {\rm with} 
\\
    n^\mu v_\mu &= 0
\,,
\quad
	\gamma ({\bm v}) 
    := 
    - n^\mu u_\mu  = \left( 1 - v^\mu v_\mu \right)^{-1/2} 
\, ,
\end{aligned}
\end{equation}
where $\bm v = v^i \bm \p_i$ is the \textit{covariant} spatial velocity field, as~opposed to the fluid's coordinate velocity field ${\bm V} =V^i \bm \p_i$
(henceforth referred to as \textit{coordinate velocity} for simplicity): 
\begin{equation}\label{eq: coord vel}
    \bm V := \ddt{\bm x}
\ ,
\quad
    n_\mu V^\mu = 0
\ .
\end{equation}

We associate the former with the spatial velocity of the spaceship with its warp field, contrary to the current literature on warp drives where \eqref{eq: coord vel} is associated with the velocity of the spaceship that carries along a prescribed warp field (see, e.g.,~\cite{Alcubierre_1994_warp, Natario_2002_warp,Visser_2021_generic_warp}).
$\bm v$ is a measure of tilt between the normal frames and the fluid frames, 
also measured through the \textit{covariant Lorentz-factor} 
$\gamma ({\bm v})$\footnote{By \textit{covariant Lorentz-factor} we emphasize that it is a function of the covariant velocity $\bm v$.}. Velocity is $\bm v$, hence $\bm u$ can be written as follows:
\begin{align}
	\bm v 
    &= 
    \frac{1}{N} \left( \bm N + \bm V \right)
\ ,
\label{eq:relat_vel}
\\
    \bm u 
    &= 
    \frac{\gamma ({\bm v})}{N} \left( N \bm n + \bm N + \bm V \right) 
\,,\, 
{\rm with}
\label{eq:four_vel_II}
\\
	\frac{\gamma({\bm v})}{N} 
 &= \left[ N^2 - (N^i + V^i) (N_i + V_i) \right]^{-1/2}
,
\end{align}
and we obtain from (\ref{eq:four_vel_II}) the component expressions for $\bm u$ and $\underline{\bm u}$ (cf.~\cite{buchertetal:generalfluid, smarr:foliation,alcub:foliation,gourg:foliation}):
\begin{equation}\label{eq:comp_u}
    \bm u 
    = \frac{\gamma ({\bm v})}{N} \left( 1, \bm V \right) 
\,, \qquad
	\underline{\bm u} 
    = \frac{\gamma({\bm v})}{N} \left( - N^2 + N^i \left( N_i + V_i \right) , \, \underline{\bm N} + \underline{\bm V} \right)\, .
\end{equation}

An important concept is that of \textit{coordinate observers}, i.e.,~those observers that keep their coordinate identity during evolution 
(see~\cite{smarr:foliation} and \cref{fig:mainfig} below). 
While in SR such observers are associated to a coordinate system of a global vector space, a~coordinate observer in GR can only be defined locally. 
The vector field $\bm \p_t$ (also called \textit{time vector}, cf., e.g.,~\cite{gourg:foliation,alcub:foliation}) is tangent to the congruence of coordinate observers' worldlines which, \textit{a priori} can be null, space- or time-like, and~characterizes the lapse function and the shift vector field introduced above, i.e.,
\begin{equation}
\label{eq: time vector}
\bm{\partial}_t = N \bm n + \bm N
\,,
\quad
    N = - n_\mu \, (\bm \p_t)^\mu
\,,
\quad
    N_i = h_{ij} (\bm \p_t)^j
\,.
\end{equation}

However, a~\textit{coordinate observer} is the one that has a normalized time-like $4$-velocity $\hat{\bm{\partial}}_t$, tangent to the observer's worldline (cf.~\cite{smarr:foliation}), 
i.e.,\footnote{Sometimes, the~\textit{coordinate observer} is associated with the \textit{time vector} (e.g.,~in~\cite{smarr:foliation}). We shall reserve the word \textit{coordinate observer} for the $4$-vector $\hat{\bm{\partial}}_t$, since by definition an \textit{observer} moves along a future-directed time-like curve with normalized $4$-velocity (cf., e.g.,~\cite{1977_Sachs_WU_MathGR} (Chapter 2)).}

\beq
\label{Hansvelocity}
    \hat{\bm{\partial}}_t 
    :=
    \left| N^2 - N^k N_k \right|^{-1/2} \bm{\partial}_t
\ \ , \ \
g_\mnu (\hat{\bm{\partial}}_t)^\mu (\hat{\bm{\partial}}_t)^\nu = -1
\ .
\eeq
\subsection{Properties of Restricted Warp~Drives}
\label{SR}

The first item in the list of restrictions imposed in the current literature, which we denote by ${\rm\bf R1}$,
is the assumption of \textit{flow-orthogonality}, i.e.,~the four-velocity $\bm{u}$ is assumed to follow the normal congruence defined by $\bm n$. 
From \eqref{eq:four_vel}, proportionality of $\bm u$ and $\bm n$ implies $\bm v = {\bf 0}$, $\gamma (\bm 0) = 1$, hence we have no tilt and ${\bm u} = {\bm n}$. 
In general, however, the~
four-velocity can be tilted with respect to the normal, a~fact that we identify below as essential to describe covariant spatial motion. 
Flow-orthogonality, ${\rm\bf R1}$, also results in the covariant restriction to irrotational flows, $\boldsymbol{\omega} = {\bf 0}$, in~components (using squared brackets for antisymmetrization):
\beq
\label{4-vorticity}
    \omega_\mnu 
    := 
    \tensor{b}{^\alpha_\nu} \tensor{b}{^\beta_\mu} \nabla_{[ \alpha} u_{\beta ]} 
    = 
    \tensor{h}{^\alpha_\nu} \tensor{h}{^\beta_\mu} \nabla_{[ \alpha} n_{\beta ]} = 0 \ ,
\eeq
where in the second equality we make use of the vanishing of the tilt, $\bm u = \bm n$, which implies 
that the projector onto the fluid rest frames $b_\mnu$ coincides with the projector onto the spatial leaves, $b_\mnu := g_\mnu + u_\mu u_\nu = h_\mnu$.

Current warp drives are based on the spatial coordinate velocity \eqref{eq: coord vel}, denoted by ${\bm V} = {\bm V}_S (t,{\bm x})$, 
where $\bm x$ locates the spaceship $S$ moving along the normal congruence $\bm n$ measured by the coordinate observer moving along $\hat{\bm \p_t}$ at each~hypersurface.

These kinematical consequences of ${\rm\bf R1}$ already indicate that this setting must be highly restrictive in the context of describing spatial motion in GR, and~this might already cause unphysical~properties.

As a further item in the list of restrictions, named ${\rm\bf R2}$, the \textit{lapse function} and \textit{shift vector} are \textit{a priori} specified. 
Current warp drive concepts are based on the shift vector to describe spatial motion, specified to comply with ${\rm\bf R1}$. 
The shift vector determines the relation between the Eulerian\footnote{In the covariant setting, observers moving along the normal congruence are commonly named \textit{Eulerian observers}.} and coordinate observers at any point on the \{$t = \const$\} hypersurfaces (cf.~\eqref{eq: time vector}). In~particular, from~${\rm\bf R1}$, we have $\bm v_S = {\bf 0}$, and hence by \eqref{eq:relat_vel} we find ${\bm N} = - {\bm V}_S = - V^i_S \, \bm \p_i$, so that the spaceship, which in this case coincides with the Eulerian observer, is shifted by $- {\bm V}_S$ to compensate for the tilt, resulting in a flow-orthogonal~foliation.

In most cases the lapse is assumed to be (up to a constant) $N=1$.  
Combining this assumption with ${\rm\bf R1}$, we speak of a \textit{geodesic slicing} of spacetime (cf.~\cite{gourg:foliation, smarr:foliation}), since the Eulerian observers are freely falling, i.e.,~the four-acceleration $\bm a$ then vanishes:
\beq
a^\mu = u^\alpha \nabla_\alpha u^\mu = n^\alpha \nabla_\alpha n^\mu = 
h^{\mu \nu} D_\nu \ln N = 0 \ ,
\eeq 
where we denote the covariant derivatives associated to $\bf g$ and $\bf h$ by $\nabla_\alpha$ and $D_\alpha$, respectively.  
In the second equality, we again made use of the vanishing of the tilt.
For the third equality see~\cite{gourg:foliation, buchertetal:generalfluid, smarr:foliation}.

In the following, we look at the proper time intervals, defined through the real parameter $\tau$, ${\mathrm d}\tau^2 : = - {\mathrm d}s^2$. We distinguish the proper time intervals for 
the coordinate observer ${\mathrm d}\hat\tau$ and for the Eulerian observer (here also the spaceship) ${\mathrm d}\tau_E = \diff \tau_S$.
According to the line-element \eqref{eq: ADM metric}, 
the choices $N=1$ and ${\bm N} = -{\bm V}_S$ imply ${\mathrm d}\tau_E  =  {\mathrm d}t$, i.e.,~there is no time dilatation between the clock of the spaceship and the coordinate time.
However, there is time dilatation between the clock of the coordinate observer with four-velocity $\hat{\bm{\partial}}_t$ and that 
in the spaceship. From~\eqref{eq: ADM metric}, we obtain 
with $\hat{\bm V} = \diff \hat{\bm x}/\diff t = {\bf 0}$ for the coordinate observers at their fixed position $\hat{\bm x}$, and~$\diff t = \diff \tau_E$:
\beq\label{timedilatHans}
    \diff \hat{\tau} = 
    \sqrt{1 - h_{ij} V_S^i V_S^j} \,\diff t 
    =:
    \Gamma^{-1}({\bm V}_S) \,\diff \tau_E\ ,
\eeq
where we introduce the \emph{coordinate Lorentz factor} $\Gamma ({\bm V})$ as a function of the coordinate velocity to distinguish it from the \emph{covariant Lorentz factor} $\gamma (\bm v)$\footnote{In~\cite{Alcubierre_1994_warp}, Alcubierre compares the proper time of the spaceship with a ``distant observer in the flat region'', i.e.,~an observer at rest at infinity since the Alcubierre metric is asymptotically flat. For~such observer ${\bm N}_{\infty} = - {\bm V}_{\infty} = {\bf 0}$ and $N = 1$, hence $\hat{\bm \p}_t = \bm \p_t = \bm n$. Therefore, there is no time dilatation between the spaceship and the distant observer. However, there is time dilatation between the spaceship and an observer in the vicinity of the spaceship. More precisely, since the shift vector is assumed to rapidly tend to zero outside the `warp bubble', the~coordinate observer becomes an Eulerian observer and time dilatation vanishes. But, in~other models where the shift vector decays more slowly there is more significant time dilation.}.

In conclusion, for~\Rwarp{} models, \eqref{eq:comp_u} reduces to the following components of the four-velocity of the spaceship: 
${\bm u}_S = (1,{\bm V}_S)$ and $\underline{\bm u}_S = (-1,\bm 0)$.

The final restriction adopted for most of the \Rwarp{} models\footnote{With some exceptions including the metric introduced by Van Den Broeck~\cite{Broeck_1999_warp}, where the author generalized the Alcubierre metric slightly by considering conformally flat slices.} that we mention in this paper is the restriction to \textit{flat spatial hypersurfaces}, ${\rm \bf R3}$. As~we found and eventually communicate in a forthcoming work, this implies a very narrow class of possible motions and stress-energy sources when imposing the Einstein equations. A~comprehensive discussion of the main restrictions ${\rm \bf R1}$, ${\rm \bf R2}$, and ${\rm \bf R3}$ and further employed restrictions will be provided in forthcoming~works.

Let us illustrate with the example of the Alcubierre spacetime some immediate consequences of the imposed restrictions.
Using one of the Einstein equations, the~energy constraint (assuming  now ${\rm \bf R1}$) is
\beq
    \CR = 16 \pi G \epsilon + 2\Lambda - 2\II (\bm \theta) 
\,,
\eeq
where $\CR$ is the scalar curvature of the space sections, $\epsilon$ the energy density, $\Lambda$ the cosmological constant, and~$\II (\bm\theta)$ the second principal invariant of the expansion tensor $\theta_{\mnu} := \tensor{b}{^\alpha_\mu} \tensor{b}{^\beta_\nu} \nabla_{( \alpha} u_{\beta )}=\tensor{h}{^\alpha_\mu} \tensor{h}{^\beta_\nu} \nabla_{( \alpha} n_{\beta )}$. We have the additional restriction ${\rm \bf R3}$, $\CR=0$, so that the expansion tensor reduces to $\theta_{ij}= \p_{(i} V_{j)}$, and~therefore $2\II (\boldsymbol{\partial}{\bm V})= 2\II (\bm\theta)+ 2\boldsymbol{\Omega}^2 = 
\boldsymbol{\partial}\cdot [{\bm V}(\boldsymbol{\partial}\cdot {\bm V})-
({\bm V}\cdot\boldsymbol{\partial}){\bm V}]$ (with $\boldsymbol{\partial}$ denoting the nabla operator, $\bm \p \bm V := \p_i V_j \diffb x^i \otimes \diffb x^j$, and~$\boldsymbol{\Omega}:= \frac{1}{2} \, \bm \p \times {\bm V}$). Assuming now ${\rm \bf R2}$,  together with Alcubierre's velocity model, $V^i = V\delta^i_{\; 1}$, we find
$\II (\boldsymbol{\partial}{\bm V}) =0$, and~we are left with $8\pi G \epsilon + \Lambda = - \boldsymbol{\Omega}^2 = -[(\p_2 V)^2 + (\p_3 V)^2]/4$. 
We conclude that for irrotational Alcubierre warp drive, $\boldsymbol{\Omega}={\bf 0}$, (i) for $\Lambda = 0$, the~energy density has to vanish, and (ii) for $\Lambda \ne 0$, the~energy density has to compensate the cosmological constant in the form of a negative `vacuum' density $\epsilon = -\Lambda / 8\pi G$. 
We notice in particular that the problem of negative energy density---often discussed in the literature---disappears for $\Lambda \le 0$ in the case of vanishing coordinate vorticity. This shows the importance of $\bm \Omega$ for \Rwarp{} models, although~only a few \Rwarp{} models assume that $\bm \Omega$ vanishes.

\subsection{Properties of Tilted Warp~Drives}
\label{S}

In general, the four-velocity $\bm u$ is not aligned with the normal $\bm n$ of the foliation. With~non-vanishing tilt, there is a physical spatial displacement of the spaceship $S$ away from the normal congruence of a left-behind Eulerian observer, denoted by the infinitesimal vector $\diff\boldsymbol{\ell}_S$. It obeys via vector addition, from~which \eqref{eq:four_vel} and \eqref{eq:relat_vel} naturally follow\footnote{The relation \eqref{triangle} might merely serve as an intuition as there is no need to define the spatial covariant velocity in this way because \eqref{eq:four_vel} is covariantly well-defined (see, e.g.,~\cite{alcub:foliation} (Section 7.3)). The~reader may consult \cite{gourg:foliation} (Section 6.3.1), in~particular Figures~6.1 and 6.2. (cf. Figure~\ref{fig:mainfig}b below). We here think of the covariant spatial velocity ${\bm v}_S$ that results in the infinitesimal displacement vector $\diff \bm \ell_S = {\bm v}_S \diff\tau_E$ after the elapsed proper time differential $\diff\tau_E$ (which are both not exact forms, although~we use the same symbol $\diff$ by abuse of notation; this also applies to \eqref{timedilatHans}).}:
\begin{equation}
\label{triangle}
\diff\boldsymbol{\ell}_S = {\bm u}_S \, \diff\tau_S - {\bm n} \, \diff\tau_E = {\bm N} {\mathrm d}t+ {\mathrm d}{\bm x} 
\ \ , \ \  {\bm v}_S = \frac{\diff\boldsymbol{\ell}_S}{\diff\tau_E} \ , 
\end{equation}
where the second equality in the first equation gives its coordinate representation. 
We denote by $\tau_E$ and $\tau_S$ the proper times measured along the normal and along the worldline of the spaceship, respectively. 
Recall that for the restricted warp drive concepts of Section~\ref{SR}, $\diff\boldsymbol{\ell}_S = {\bf 0}$. 
Dividing \eqref{triangle} by $\diff\tau_E$, we find in comparison with \eqref{eq:four_vel}, 
$\diff\tau_E = \gamma ({\bm v}_S)\, \diff\tau_S$, 
or more directly, from~the very definition of $\bm u$ and \eqref{eq:four_vel}, 
\begin{equation}\label{timedilat2}
\begin{aligned}
    \gamma (\bm v_S)
    &=
    - n_\mu u^\mu 
    =
    N u^0
    =
    N \frac{\diff t}{\diff \tau_S}
    =
    \frac{\diff \tau_E}{\diff \tau_S}
\\
    &\Rightarrow \ \diff\tau_E = \gamma ({\bm v}_S)\, \diff\tau_S \ ,
\end{aligned}
\end{equation}
showing that for tilted warp drives there is time dilatation between the left-behind Eulerian observer $E$ in the normal congruence and the spaceship $S$ measured by the tilt.  
Since $\gamma (\bm v_S) > 1$, the~proper time interval in the tilted rest frames of the spaceship is smaller compared to the proper time interval of an observer moving along the normal congruence. In~other words, the~clock in the spaceship runs slower compared with the Eulerian observer's~clock.

We now introduce a particular realization of tilted warp drives, named \textit{Lagrangian} \Twarp{}, in~order to have a concrete setting that we can easily compare with \Rwarp{}. For~this, we employ a Lagrangian coordinate system that is centered on the spaceship, $ \bm x \mapsto {\bm X} = (X^i)$, with~${\mathrm d}{\bm X}/{\mathrm d}t = {\bf 0}$. Hence, the~coordinate velocity measured in the rest frame of this spaceship vanishes, ${\bm V}_{S} (t,{\bm X}) := {\mathrm d}{\bm X}/{\mathrm d}t = {\bf 0}$.

We furthermore choose\footnote{We herewith also specify a relation between lapse and shift 
\textit{a priori}. We believe, however, that this choice implies a number of advantages, at~least serving as a first step to realize the \Twarp{} concept with sufficient generality.} lapse and shift such that the
four-velocity of the spaceship attains the Lagrangian form, ${\bm u}_S = (1, \bm 0)$ (cf.~\cite{friedrich, smarr:foliation}), leading via \eqref{eq:comp_u} to
$N^2 - N^k N_k = 1$, $N=\gamma (\bm v_S)$. For~this choice, we also have ${\mathrm d}\tau_{S} = {\mathrm d}t$ and hence $N=\gamma ({\bm v}_{S}) > 1$ and $\bm N = \gamma ({\bm v}_{S}) {\bm v}_{S}$.
In the tilted Lagrangian description, the~four-velocities of the spaceship and that of the coordinate observer (and the time vector) thus all coincide, ${\bm u}_S = \hat{\boldsymbol{\partial}}_t = \boldsymbol{\partial}_t$.

In conclusion, for~the special case of Lagrangian \Twarp{}, \eqref{eq:comp_u} reduces to the following components of the four-velocity of the spaceship:
${\bm u}_S = (1, \bm 0)$ and $\underline{\bm u}_S = (-1,\gamma ({\bm v}_S) \underline{\bm v}_S)$.

One advantage of the Lagrangian choice (out of many others) is the possibility of comparison between the spaceship of Section~\ref{SR} moving along the normal congruence and~the tilted spaceship of Section~\ref{S} moving away from Eulerian observer but remaining within the same spatial leaves parametrized by $t$.

\subsection{Comparison of Restricted and Tilted Lagrangian Warp~Drives}
\label{comparison}

\begin{figure}[H]
 \centering
  \subfloat[Flow-orthogonal (coordinate velocity).]{
    \includegraphics[width=0.9\textwidth]{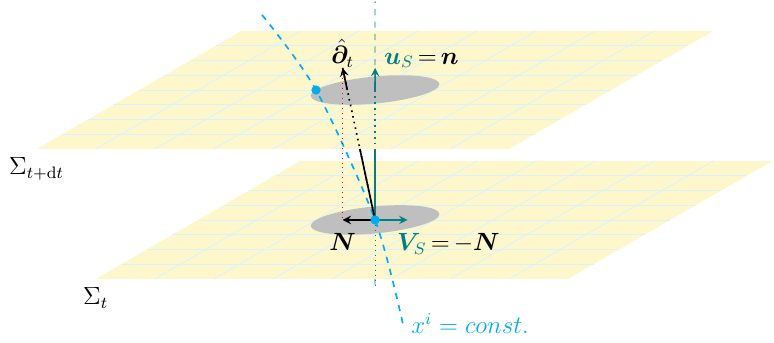}
    \label{fig: flow-orthogonal}
  }
  \hspace{0.05\textwidth}
  \subfloat[Lagrangian tilted (covariant velocity).]{
    \includegraphics[width=0.9\textwidth]{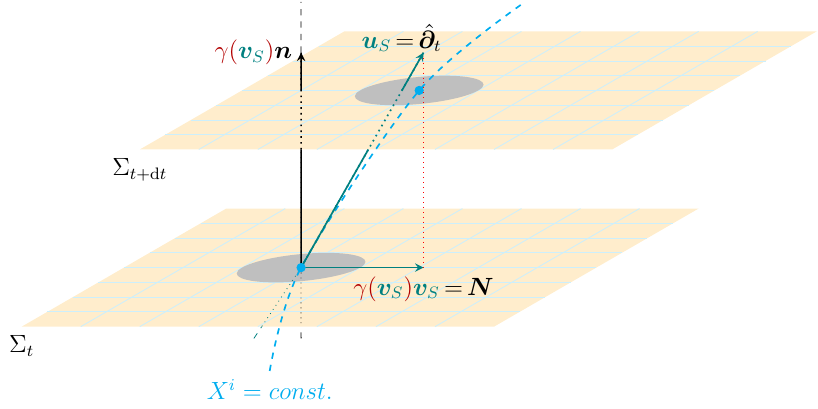}
    \label{fig: tilted}
  }
  \caption{Schematic comparison of the spacetime architecture for (\textbf{a}) Alcubierre and Nat\'ario \Rwarp{} and~(\textbf{b}) Lagrangian \Twarp{}.}
  \label{fig:mainfig}
\end{figure}

\cref{fig:mainfig}
compares the spacetime architecture of current warp drive proposals with four-velocity, ${\bm u}_S = (1,{\bm V}_S)$ and $\underline{\bm u}_S = (-1,\bm 0)$, with~the covariant proposal advanced in this paper  and specified to a Lagrangian four-velocity, ${\bm u}_S = (1, \bm 0)$ and $\underline{\bm u}_S = (-1,\gamma ({\bm v}_S) \underline{\bm v}_S)$, while considering flat spatial slices according to ${\rm \bf R3}$ in both cases for the purpose of comparison (to convey this, we depict flat slices for illustration only).

We compare the two situations in Figure~\ref{fig:mainfig}. \cref{fig:mainfig}a shows two consecutive spatial slices at a lapse distance $N=1$. A~coordinate observer keeps its coordinate identity along its tangent vector field $\hat{\boldsymbol{\partial}}_t$. From~this shifted observer, the `warp bubble' is observed to move with velocity ${\bm V}_S$ that compensates the coordinate shift ${\bm N} = - {\bm V}_S$. There is no physical covariant motion of the `warp bubble' that evolves in time along the normal to the hypersurfaces, ${\bm u}_S = {\bm n}$, hence it does not leave the normal congruence set by the spacetime foliation. Recall that, for~the typical description of the warp field in terms of a `warp bubble' with rapidly decaying shift outside of the bubble wall,   $\hat{\boldsymbol{\partial}}_t$ rapidly becomes an Eulerian observer away from the warp~field. 

On the contrary, in~Figure~\ref{fig:mainfig}b, the four-velocity ${\bm u}_S$ of the spaceship is tilted with respect to the normal vector $\bm n$. 
This implies a non-vanishing \textit{proper velocity}
$\gamma ({\bm v}_S) {\bm v}_S = {\mathrm d}{\bm\ell}_S / {\mathrm d}\tau_S$ which is the projection of $\bm u$  onto the hypersurface $\Sigma_t$, 
$\tensor{h}{^\mu_\nu} u^\nu = \gamma({\bm v}_S) v_S^\mu$ (cf.~\eqref{eq:four_vel}).  
Lapse and shift are chosen such that the spaceship is described in a Lagrangian frame that moves with it: $N^2 - N^k N_k = 1$, $N=\gamma (\bm v_S)$.  
Therefore, $\diff \tau_S = \diff t$ (see Section~\ref{S}). 
Recall that for \Rwarp{} models $N = 1 = \gamma (\bm 0)$; as a result, we also have $\diff \tau_S = \diff \tau_E = \diff t$. But this setting does not involve a non-vanishing covariant~velocity.

In this tilted Lagrangian case, a~traveler within the spaceship has coordinate velocity ${\bm V}_{S} = {\bf 0}$, thus keeping its coordinate identity as the coordinate observer in Figure~\ref{fig:mainfig}a, but~here the vector field $\hat{\boldsymbol{\partial}}_t$ coincides with the four-velocity of the spaceship ${\bm u}_S$. 
This has the further advantage that the spaceship is described intrinsically, while in the restricted case the description remains~extrinsic. 

In both cases, the respective proper times equal the coordinate time labeling the spatial leaves ($\diff \tau_S = \diff t$). 
Both figures are extrinsic in the sense that we compare the two situations within the depicted normal congruence. However, the~proper time intervals of the Eulerian observers in Figure~\ref{fig:mainfig}a,b are different: for the restricted case, ${\mathrm d}\tau_E = {\mathrm d}t$, but~for the tilted cases according to \eqref{timedilat2}, ${\mathrm d}\tau_E = \gamma (\bm v_S) \,{\mathrm d}t$.

\section{T-Warp: A New Concept Beyond Current~Proposals}
\label{outlook}

Here, we provide a simple example of a \Twarp{} which generalizes the \natario{} warp drive to the tilted case. Hence, we consider flat spatial slices, i.e.,~$h_{ij} = \delta_{ij}$. However, we use the general relation \eqref{eq:relat_vel} to have $\bm N = N \bm v - \bm V$ and allow for non-constant lapse function $N = \gamma(\bm v) \geq 1$. Therefore, the~line element \eqref{eq: ADM metric} takes the following form:
\begin{equation}\label{eq: tilted_metric}
    \diff s^2
    =
    - \left( 1 - V^2 + 2 \gamma V^i v_i \right) \diff t^2 
    +
    2 (\gamma v_i - V_i) \diff t \diff x^i 
    +
    \delta_{ij} \diff x^i \diff x^j
\,,
\end{equation}
which generalizes the \natario{} metric to a \Twarp{}; indeed, if~$\bm v \rightarrow \bm 0$ (hence $\gamma(\bm v) \rightarrow 1$), one recovers the \natario{} metric. 
Put differently, when the spaceship follows a background flow with flat spatial slices, moving along a designated unit normal vector $\bm{n}$, the~\natario{} metric is recovered. However, if~the pilot opts for a different tilted trajectory, the~metric \eqref{eq: tilted_metric} arises. Moreover, if~we consider a comoving frame, we then have $V^i = 0$, which gives rise to the Lagrangian description explained in \Cref{S} (cf.~\cref{fig:mainfig}b). A~quick examination shows that both Eulerian and fluid energy densities can generally be non-negative. However, we do not necessarily conclude that these models are free of issues related to energy conditions, which requires more careful analysis. Instead, it suggests that altering the kinematic perspective entails additional~possibilities. 

It is evident from our discussion and illustrations that relaxing the imposed restrictions, ${\rm\bf R1}$, ${\rm\bf R2}$, and ${\rm\bf R3}$, discussed in Section~\ref{SR}, 
will allow for the following improvements and suggestion of a new concept for warp drive spacetimes. 
We name this concept \Twarp{} for \textbf{T}ilted warp drive as opposed to \textbf{R}estricted warp drive, \textbf{R}-Warp, that obeys \textit{all} of the above restrictions.  
\Twarp{} implies that the restrictions imposed by \textbf{R}-Warp
are all relaxed. Tilt is a hierarchically superordered property that subsequently implies,
from the existence of a non-vanishing velocity $\bm v$, non-vanishing  acceleration $\bm a$, vorticity $\boldsymbol{\omega}$, and spatial curvature in general, while for \Rwarp{} these fields all vanish.
Subclasses in which one or more of the physics of \Twarp{} are neglected can still be~considered.

The spatially projected four-acceleration will replace the \textit{coordinate  spatial acceleration}, ${\bm A}:= \diff \bm V/
\diff t$, and~the spatially projected covariant vorticity will replace the \textit{coordinate spatial vorticity} $\boldsymbol{\Omega}= \frac{1}{2} \bm \p \times {\bm V}$ of the \textbf{R}-Warp~concept.

For \Twarp{}, the covariant Lorentz factor becomes singular at the time when the tilted four-velocity becomes light-like. 
Although the spaceship itself is not affected by the extrinsic observations of the left-behind Eulerian observer, the~description using the covariant spatial velocity breaks down.
Note, however, that the value of the Lorentz factor or of the tilt depends on the relative angle between $\bm u$ and $\bm n$ at the location of the moving~spaceship.

In GR, we may not necessarily think of large velocities in terms of $\bm v$, as~the cosmological example of expanding space, described within a flow-orthogonal setting, shows. Thus, flow-orthogonality is not \textit{per se} excluding covariant warp effects. It makes physical sense if some restrictions of the \Rwarp{} concept are removed, i.e.,~keeping ${\rm\bf R1}$, adopting a Lagrangian description for ${\rm\bf R2}$, i.e.,~using Gaussian normal coordinates ($N = 1 =\gamma (\bm 0)$, ${\bm N}={\bf 0}$), 
but necessarily relaxing ${\rm\bf R3}$; a flow-orthogonal and covariant warp concept that we may call Lagrangian \Fwarp{} arises. As~an example, we mention the possibility to describe propagating gravitational waves in a Lagrangian flow-orthogonal setting~\cite{2017_AlRoumi_Buchert_Wiegand_LGW}. It is an indirect consequence of the equivalence principle allowing to exchange the roles of curvature and acceleration. The~restrictions to irrotational and non-accelerating flows, however, remain. Their relaxation, i.e.,~moving from \Fwarp{} to \Twarp{}, may be crucial for warp realizations, as~the example below~illustrates.

We give an example for interesting new features of \Twarp{}.
As \Twarp{} allows for acceleration and vorticity, let us quote a relation relating them, taken from~\cite{buchertetal:generalfluid}: $\omega_{\mu \nu} = u_{[\mu} a_{\nu]} + \partial_{[\nu} u_{\mu]}$. 
In Lagrangian coordinates, $a_i = {\mathrm d}u_i / {\mathrm d}t = {\mathrm d}u_i/{\mathrm d}\tau_S$.
Combining this property with $a_0 = 0$, from~$a_\alpha u^\alpha = 0$, we can thus write the spatial components of the vorticity, $\omega_{\mu\nu}u^\nu = 0$, as~$\omega_{i j}=u_{[i}a_{j]}+\p_{[j} u_{i]}$, 
which shows that vorticity and acceleration are mutually dependent. This relation can be read such that acceleration contributes to vorticity, but~in the present context it can also be read such that active generation of vorticity by the spaceship can produce~acceleration.

Spatial curvature leads to the image of a spaceship that not only leaves a left-behind Eulerian observer but~also leaves the spatial tangent space of the Eulerian observer to subsequent tangent spaces with tilted unit normal vectors.
It is to be analyzed whether this observer will eventually experience the travel of the spaceship as an accelerated motion in space with its subsequent disappearance behind a horizon at a time that depends on acceleration and curvature. A~warp concept that assumes that the spaceship is capable of intrinsically warping the space around its rest frames through the correspondence between
covariant acceleration (e.g., induced by tilt and covariant vorticity) and spatial curvature would add a truly general-relativistic~element.

We think that the presence of tilt for a physical model of warp is essential.
To clarify this point, one can think of a warp drive that traverses interplanetary, interstellar, or intergalactic distances. The~normal congruence would be associated with the frame of the solar system, the~galactic plane, or the frame of a comoving cosmological background, respectively. In~all cases, there is a background gravitational field against which a warp drive should act, e.g.,~to escape the Sun's gravitational field (note that spatial curvature is important in all cases even if metric perturbations are very small~\cite{2009_Buchert_Elst_Ellis_magnitude}). For \Rwarp{},
it is assumed that the background is the Minkowski spacetime 
and the corresponding warp drive is roving around in a vacuum space, which---as we show in \cref{fig:mainfig}a---represents no physical motion. As~a result, a~physical proposal for warp drive, we believe, should be based on a covariant formalism and must involve tilted motion if spatial acceleration and vorticity are~non-vanishing.

For the understanding of the notion of the \textit{warp field} or `warp bubble', we may consider the design for the \Rwarp{} concept, where a form function of the bubble comoving with the spaceship may be given as suggested by Alcubierre~\cite{Alcubierre_1994_warp}. The~warp field is treated as being part of a continuum, since a velocity field ${\bm V}_S (t,{\bm x})$ is given and the expansion profile is calculated from it. If~the warp field were to be a solution of the Einstein equation, initial data would have to be prescribed and the `warp bubble' would then be deformed in the course of evolution. 
For the common realizations of \Rwarp{}, the~warp field is assumed to be unaffected along the trajectory of the spaceship.
For the Lagrangian \Twarp{}, given initial data evolves with a change in the morphology of the Lagrangian boundary of the `warp bubble', described as a compact spatial domain. In~this latter case, the morphological evolution can be linked to the motion itself \cite{2008_Buchert_status_report} (Section 3.1.2\textit{ff}). 
Hence, as~for the \textit{matter model}, we adopt another, more conservative point of view: contrary to the body of the literature, we here emphasize considering the Einstein equation from `right to left', i.e.,~first specifying a physically reasonable matter model and then solving the Einstein equation, as~emphasized in~\cite{1971_Synge_GR, 2012_Ellis_Maartens_MacCallum_RelCos} (for a recent account see~\cite{2024_Ellis_Garfinkle_G-method}). The~usual approach, e.g.,~for the Alcubierre warp drive, is that one runs the Einstein equation from `left to right' (the so-called Synge 
\textit{G-method} \cite{1971_Synge_GR,2024_Ellis_Garfinkle_G-method}), hence eventually ending up with an unphysical matter model, named \textit{exotic}, bearing several detriments including violation of (some) energy conditions (cf.~\cite{2024_Ellis_Garfinkle_G-method}). Our proposal will certainly shrink the set of mathematically possible solutions to the Einstein equation. Nevertheless, the~outcome of the intended approach may open doors to construct physical warp~drives.

In forthcoming work, we will employ the Einstein equation to dive deeper into the consequences of the commonly imposed restrictions and extend our proposal of \Twarp{}. By~relaxing those restrictions, we think that there is much room for resolving previously discussed problems, e.g.,~the need for exotic energy densities. These problems will appear in a different light within the proposed~setting.\\


\funding{This work is a spin-out from results of a project that has received funding from the European Research Council (ERC) under the European Union's Horizon 2020 research and innovation programme (grant agreement ERC
advanced grant 740021-ARTHUS, PI: TB). }

\acknowledgments{We thank Doris Folini for valuable remarks on the manuscript, Asta Heinesen, Jan Ostrowski, Nezihe Uzun for useful~discussions.}

\bigskip
\reftitle{References}

\end{document}